\documentstyle[twocolumn,aps,epsf,amsbsy]{revtex}

\newcommand{\be}{\begin{equation}}
\newcommand{\ee}{\end{equation}}
\newcommand{\Past}        {  \stackrel{\leftarrow} {S} }
\newcommand{\past}        {  \stackrel{\leftarrow} {s} }
\newcommand{\Future}      {  \stackrel{\rightarrow}{S} }
\newcommand{\future}      {  \stackrel{\rightarrow}{s} }

\newcommand{\PastBlock}   { {\stackrel{\leftarrow} {S}}^L }

\newcommand{\FutureBlock} { {\stackrel{\rightarrow}{S}}^L }

\newcommand{\hmu}         { {h_\mu} }
\newcommand{\PastRate}    { {\stackrel{\leftarrow} {h}}_\mu }
\newcommand{\FutureRate}  { {\stackrel{\rightarrow}{h}}_\mu }
\newcommand{\InternalRate}{ h_\mu^{\cal X} }
\newcommand{\ObserveRate} { h_\mu^{\cal A} }
\newcommand{\AllPasts}    {  \stackrel{\leftarrow} {\rm {\bf S}} }
\newcommand{\CausalStateSet}    { \boldsymbol{\cal S} }
\newcommand{\CausalStateRate}   { h_\mu^{\CausalStateSet} }
\newcommand{\AlternateStateSet} { \boldsymbol{\cal R} }

\begin{document}
\draft

\title{Thermodynamic Depth of Causal States:\\
When Paddling around in Occam's Pool\\
Shallowness Is a Virtue}

\author{James P. Crutchfield}
\address{Physics Department, University of California, Berkeley, CA 94720-7300\\
and the Santa Fe Institute, 1399 Hyde Park Road, Santa Fe, NM 87501\\
Electronic address: chaos@santafe.edu}

\author{Cosma Rohilla Shalizi}
\address{Physics Department, University of Wisconsin, Madison, WI 53706\\
and the Santa Fe Institute, 1399 Hyde Park Road, Santa Fe, NM 87501\\
Electronic address: shalizi@santafe.edu}

\date{\today}
\maketitle

\bibliographystyle{unsrt}

\begin{abstract}
Thermodynamic depth is an appealing but flawed structural complexity measure.
It depends on a set of macroscopic states for a system, but neither its original
introduction by Lloyd and Pagels nor any follow-up work has considered how to
select these states.  Depth, therefore, is at root arbitrary. Computational
mechanics, an alternative approach to structural complexity, provides a
definition for a system's minimal, necessary causal states and a procedure
for finding them. We show that the rate of increase in thermodynamic depth,
or {\it dive}, is the system's reverse-time Shannon entropy rate, and so depth
only measures degrees of macroscopic randomness, not structure. To fix this
we redefine the depth in terms of the causal state
representation---$\epsilon$-machines---and show that this representation gives
the minimum dive consistent with accurate prediction. Thus, $\epsilon$-machines
are optimally shallow.
\end{abstract}



\vspace{-0.25in}
\tableofcontents


\section{Natural Complexity}

Dissipative dynamics, symmetry breaking, phase transitions, bifurcations, and
pattern formation, acting over different temporal and spatial scales, at
different levels and on different substrates, are presumably responsible for
assembling and freezing in the wide diversity of structures observed in the
natural world.  Each of these processes has its more-or-less well-developed
foundations.  But where are the principles that define and describe their
products?  What is structure itself?  Does each and every particular
combination of forces lead to a different and unique class of natural
structure, requiring its own vocabulary and theory?  And, how do we detect that
some new structure has emerged in the first place?

These and related questions about nature's complexity have engaged a large
number of researchers for several decades now; for a sampling see e.g. Refs.
\cite{Gras86,Lind88a,Wack94,Feld97,Badi97} and references therein. One focus
has been on quantitative measures of the complexity of natural objects and
of the processes
that bring them into existence---measures that capture properties more
interesting than mere randomness and disorder.  Existing theory, such as is
found in statistical mechanics, provides relatively well-understood measures of
disorder in (say) temperature and thermodynamic entropy, and of the flow of
energy that can do work in the various free energies.  While many applications
and problems remain, there is little pressing need for new conceptual
approaches to randomness and energy transduction.  However, when it comes to
structure something is missing---something else must be invented and then added
to physical theory to account for, work with, and quantify kinds of structure.

One class of approaches to natural complexity is based on the theory of
sequential discrete computation \cite{Hopc79,Papa94a}---the theory of how
sundry sorts of discrete-state devices process information at varying levels of
sophistication. The resulting measures of complexity ultimately express
structural properties in terms of universal Turing machines.  Unfortunately,
almost all interesting mathematical and quantitative questions about these
measures of structure inherit the uncomputability associated with those
all-powerful machines. More fundamentally, though, the idea that everything in
the world is really a discrete-state computer strikes one as inadequate; at
a minimum nature is parallel, continuous, spatially extended, noisy, and
quantum mechanical.

Fortunately, in the thermodynamic depth of Lloyd and Pagels \cite{Page88a} we
have a proposal for a noncomputation-theoretic, empirically calculable measure
of the complexity of processes.  One central motivation for defining the
thermodynamic depth is that it is small both for regular and for random processes.
Thus, one of its appealing features is that depth measures something other than
randomness---a property already well-captured by both Kolmogorov-Chaitin
complexity \cite{Kolm65,Chai66,Vita93} and Shannon entropy rate
\cite{Shan48,Kolm59,Cove91,Brud83a}.

In this brief note we introduce the required background for thermodynamic depth
\cite{Page88a} and for an alternative approach to natural complexity, called
computational mechanics \cite{Crut89,Crut92c}, that extends statistical
mechanics to address issues of structure in a direct way. We review the
definition of thermodynamic depth and apply it to several simple Markov
processes, revealing several ambiguities. To remove them we redefine
the depth in terms of a representation based on causal states, those states
through which computational mechanics views the minimal structure of a system
\cite{Crut89,Crut92c}. We then prove our main results on the predictive
optimality and minimality of the causal state representation.  Finally, we draw
a number of conclusions about using thermodynamic depth as a measure of
structural complexity in natural processes.

\section{Processes}

Following Lloyd and Pagels we focus on discrete-time processes and consider a
given process as a joint probability distribution ${\rm Pr} (\ldots, X_{-1},
X_0, X_1, \ldots)$ over random (``microscopic'') variables $X_t$ at each time
$t$ that take values $x_t$ in a continuous state space $\cal X$.  In accord
with experimental constraints, we assume that the process is not observed
directly, but states are in fact measured via a finite-precision instrument.
The result is that our description of the process is in all practicality a
joint distribution over a chain $\stackrel{\leftrightarrow}{S} ~\equiv~ \ldots
S_{-2} S_{-1} S_0 S_1 \ldots$ of discrete-valued random variables $S_t$
that range over a finite set ${\cal A}$ of observed states.  (Although our
notation differs, this setup follows the account in Ref. \cite[p. 194]{Page88a}
of ``macroscopic'', ``measured'', or ``coarse-grained'' states as partitions
of the underlying microscopic state space.)

We divide the chain into two semi-infinite halves by choosing a time $t$ as the
dividing point.  Denote the past by
\begin{equation}
\stackrel{\leftarrow}{S_t} \equiv \ldots S_{t-3} S_{t-2} S_{t-1} 
\label{FutureHalf}
\end{equation}
and the future by
\begin{equation}
\stackrel{\rightarrow}{S_t} \equiv S_t S_{t+1} S_{t+2} S_{t+3} \ldots \; .
\label{PastHalf}
\end{equation}
We will assume that the observed process is described by a temporal
shift-invariant measure $\mu$ on bi-infinite realizations $\cdots s_{-2} s_{-1}
s_0 s_1 s_2 \cdots, s_t \in {\cal A}$.
The measure $\mu$ induces a family of distributions. Let ${\rm
Pr}(s_t)$ denote the probability that at time $t$ the random variable $S_t$
takes on the particular value $s_t \in {\cal A}$ and ${\rm Pr} (s_{t+1} ,
\ldots , s_{t+L})$ the joint probability over sequences of $L$ consecutive
measurements.  Consistent with Ref. \cite{Page88a} we assume time-translation
symmetry\footnote{For the dissipative, Hamiltonian, and quantum systems
considered Ref.  \cite[pp. 194--5]{Page88a} assumes, moreover, that the
probabilities over the microscopic states are uniform with respect to Lebesgue
measure and that the probabilities of sequences over the coarse-grained state
space are time invariant.} and so ${\rm Pr}( s_{t+1},\ldots, s_{t+L})={\rm
Pr}(s_1, \ldots , s_L )$. We denote a sequence of $L$ consecutive measurements
by $S^L \equiv S_1 \ldots S_L$; when looking to the future (past) the sequence
is denoted $\FutureBlock$ ($\PastBlock$). (In dropping the time index from
Eqs. (\ref{FutureHalf}) and (\ref{PastHalf}) we implicitly take $t=0$.)
We shall follow the convention that a capital letter refers to a random
variable, while a lower case letter denotes a particular value of that
variable.  Hence, $s^L$ will denote a particular measurement sequence of
length $L$.

\section{Entropy and Randomness}

The average uncertainty of an $L$-sequence $S^L$ is given by the Shannon
entropy of the joint distribution ${\rm Pr} (S^L)$ \cite{Cove91}:
\begin{equation}
  H[ S^L ] \equiv - \sum_{s^L \in {\cal A}^L}
    {\rm Pr} (s^L) \log_2 {\rm Pr} (s^L) \;.
\label{BlockEntropy}
\end{equation}
Looking forward in time the rate of increase of this
uncertainty is defined by the entropy rate
\begin{equation}
  \FutureRate \equiv \lim_{L \rightarrow \infty} \frac{H[\FutureBlock]}{L} \;,
\label{ForwardRate}
\end{equation}
where $\mu$ denotes the above-mentioned measure.  The quantity $\FutureRate$
measures the irreducible randomness in the generation of future behavior: the
randomness that remains after the correlations over longer and longer futures
are taken into account.  The reverse-time entropy rate $\PastRate$ is defined
similarly in terms of $\PastBlock$ and measures historical randomness.  Both
can be expressed in terms of a conditional entropy: given knowledge of the
measurement history, the uncertainty in the next measurement $S_0$ is
\begin{equation}
  \FutureRate = H[S_0 | \Past] ~;
\label{CondlFutureRate}
\end{equation}
and similarly, given the future, we have
\begin{equation}
  \PastRate = H[S_{-1} | \Future] ~,
\label{CondlPastRate}
\end{equation}
where the entropy of a random variable $X$ conditioned on the value of another
random variable $Y$ is defined as $H[X|Y] \equiv H[X, Y] - H[Y]$.

\section{Thermodynamic Depth}

Lloyd and Pagels propose that the complexity of a macroscopic state $s \in
{\cal A}$ is determined by the history that led to $s$.  The motivation for
this is that ``complexity must be a function of the {\it process}---the
assembly routine---that brought the object into existence'', [emphasis theirs]
\cite[p. 187]{Page88a}; in particular, it is a ``measure of how hard it is to
put something together'' \cite[p. 189]{Page88a}. Starting from a distribution
over macroscopic state sequences, one first finds the probability of length-$L$
histories that end in state $s$:
\begin{eqnarray}
{\rm Pr} (S_{-L+1}, \ldots, & S_{-1} &, S_0 = s | s) \\
& \equiv &
\frac{ {\rm Pr} (S_{-L+1}, \ldots, S_{-1} , S_0 = s)} {{\rm Pr} (s)} ~.
\end{eqnarray}
Then the thermodynamic $L$-depth ${\cal D}_L (s)$ of state $s$ is defined by
the conditional entropy
\begin{equation}
{\cal D}_L (s) \equiv H[ S_{-L+1}, \ldots, S_{-1} , S_0 = s | s ] ~.
\end{equation}
(From here on we ignore the distinction in Ref. \cite{Page88a} between
``depth'' and ``thermodynamic depth'' by, in effect, setting Boltzmann's
constant to $1 / \ln{2}$.)  Averaging over all such states gives one the
$L$-depth ${\cal D}_L$ of the system as a whole:
\begin{equation}
{\cal D}_L \equiv \sum_{s \in {\cal A}} {\rm Pr}(s) {\cal D}_L (s) ~,
\end{equation}
or,
\begin{equation}
{\cal D}_L = H[ S_{-L+1}, \ldots, S_{-1} | S_0 ] ~,
\label{defDn}
\end{equation}
where we have used the identity $H[X,Y|X] = H[Y|X]$.  We define ${\cal D}_0 =
0$.

The back-stage intuition motivating thermodynamic depth is the following: if there is
little uncertainty about how to attain a macroscopic state and if trajectories
are confined within narrow bounds, then the macroscopic state is easy to
assemble. In this case the process leading to that state and generating those
trajectories is simple and the state is shallow.  If the historical uncertainty
is large and if a wide range of historical alternatives has been excluded, then
the process is complex and the macroscopic state is deep.  ``The thermodynamic
depth of a state $b$ is proportional to the amount of information [in bits]
needed to identify the trajectory that leads to $b$ given the information that
the system is in $b$'' \cite[p. 196]{Page88a}.

Like all statistical complexity measures, thermodynamic depth has forsworn
awarding high complexities to mere randomness. Ref. \cite{Page88a} states
that it
vanishes for completely random processes, as well as for totally ordered ones
\cite[pp. 187, 190--191]{Page88a}.  For systems satisfying the microcanonical
assumption of statistical mechanics, Lloyd and Pagels \cite[pp. 190,
194--5]{Page88a} provide another expression for the depth, as the difference
between a coarse-grained and a fine-grained thermodynamic entropy.  Using this
alternate expression, they argue that black holes \cite[p. 191]{Page88a}, gases
at thermal equilibrium \cite[p. 191]{Page88a} and salt crystals
\cite[p. 191]{Page88a} are shallow and the self-assembly of protein complexes
\cite[p. 196]{Page88a} is deep.  While it is sometimes easier to evaluate the
alternate expression than Eq. (\ref{defDn}), it is strictly equivalent to the
latter in the cases where the necessary (restrictive) conditions behind the
former hold, so we shall confine ourselves to Eq. (\ref{defDn}) in what
follows.

The total depth, $\lim_{L \rightarrow \infty} {\cal D}_L$, of a process might
as well be bottomless.  Like $L$-depth, it depends on a baseline.  That is, it
depends on the time when we judge the process to have started and on the depth
accumulated from the beginning of time until then.  At best, these choices can
be a bit tricky to figure.  Of greater physical significance, therefore, is the
asymptotic rate $v$ at which the depth increases, which we call {\it dive}:
\begin{equation}
v \equiv \lim_{L \rightarrow \infty} [ {\cal D}_L - {\cal D}_{L-1} ] ~.
\label{defv}
\end{equation}
The benefit of looking at a rate, which is not considered in Ref.
\cite{Page88a}, is that $v$ is independent of the origin of time and so allows
one to more fairly compare processes by their rate of depth generation.

We now show that $v$ is the reverse-time entropy rate. Recalling the
definition of conditional entropy, $H[Y|X] = H[X,Y] - H[X]$, Eq. (\ref{defv})
becomes
\begin{eqnarray}
v & = & \lim_{L \rightarrow \infty} \biggl[
		H[ S_{-L+1}, \ldots, S_0 ]  - H[ S_0 ] \biggr. \nonumber\\
  &   & \biggl. - H[ S_{-L+2}, \ldots, S_0] + H[S_0] \biggr] \\
  & = & \lim_{L \rightarrow \infty} \biggl[ H[ S_{-L+1}, \ldots, S_0 ]
        - H[ S_{-L+2}, \ldots, S_0] \biggr] \label{Alternate} \\
  & = & \lim_{L \rightarrow \infty}
		 H[ S_{-L+1} | S_{-L+2}, \ldots, S_0 ] \\
  & = & H[ S_{-L+1} | \Future_{-L+2} ]  = H[S_{-1}|\Future] \label{DiveCondlPastRate} \\
  & = & \PastRate ~,
\label{DivePastRate}
\end{eqnarray}
where the next-to-last step follows from time-translation invariance.

For later use note that, since $H[Y] \geq H[Y|X]$, it follows from
Eq. (\ref{DiveCondlPastRate}) and from translation invariance that
\begin{equation}
v \leq H[S_0] ~.
\label{v-bounded-by-entropy}
\end{equation}

For stationary or asymptotically stationary processes, we have $H[ S_{-L+2},
\ldots, S_0 ] = H[ S_{-L+1}, \ldots, S_{-1} ]$.  Thus, starting from
Eq. (\ref{Alternate}) we also conclude that
\begin{eqnarray}
\lim_{L \rightarrow \infty} & \biggl[ & H[ S_{-L+1}, \ldots, S_0 ]
        - H[ S_{-L+1}, \ldots, S_{-1}] \biggr] \\
  & = & \lim_{L \rightarrow \infty} H[ S_0 | S_{-L+1}, \ldots, S_{-1} ] \\
  & = & H[ S_0 | \Past ] \\
  & = & \FutureRate ~.
\end{eqnarray}
From this we see that (i) the forward-time and reverse-time entropy rates are
equal, $\PastRate = \FutureRate$, and (ii) they are the same as the dive:
$v = \hmu$. (From here on we drop the time arrows and denote a process's
entropy rate by $\hmu$.)

To summarize, we have shown that the Shannon entropy rate controls the
average rate of increase in the thermodynamic depth and that the dive is
invariant under time reversal. Recall that $\hmu$ also controls the average rate
of increase of Kolmogorov-Chaitin complexity \cite{Cove91}.  These aspects of
depth are not a surprise and are in accord with the claim that ``the average
complexity of a state must be proportional to the Shannon entropy of the set of
trajectories that experiment determines can lead to that state''
\cite[p. 190]{Page88a}.  From these elementary uses of information theoretic
identities it is clear at this point that thermodynamic depth measures nothing
other than the macroscopic randomness generated by a system.

\section{Something Rotten in the States}

The analysis of the previous section leaves us with a puzzle: How is it that
Lloyd and Pagels can state---e.g. on each of the first six pages of
Ref. \cite{Page88a}---that depth discounts for disorder and so captures
something other than randomness?

The problem, we claim, lies in their choice of states. In the illustrative
examples in Ref. \cite{Page88a} macroscopic states are selected that support the
desired properties of depth.  That is, the results and interpretations do not
follow from a direct application of the given definition of thermodynamic depth
alone; biases external to the definition are invoked.

Moreover, employing an appropriate set of macroscopic states is crucial for
obtaining a well-defined depth, since by judiciously redefining them one can
give the depth any value from 0 on up.  To see this, remember that the
depth is the conditional entropy of a sequence of states.  If there is only one
state, the depth vanishes.  If we make spurious macroscopic
distinctions---e.g. acting as though one state was really $n$ degenerate,
equiprobable states---we add a contribution to the depth that is proportional
to $\log{n}$.  And, we can keep doing this until the depth is as large as we
like.  (Cf. Ref. \cite{Page88a}'s discussion leading up to the example on page
191.)

The states of whichever dynamical system underlies the observed process are, at
least, unambiguous candidates for use in the calculation of depth, but have an
unfortunate habit of being unknown, redundant, or excessively fine-grained.
Lloyd and Pagels considered this problem by implication, discussing why, in
some particular cases, certain choices of state are better than others.  They
explain, for instance, on page 191 of Ref. \cite{Page88a} how an unfortunate
choice of measurements can make even systems in thermodynamic equilibrium quite
deep.  But they neither presented a procedure for picking sets of states nor
gave general criteria for ranking possible alternative selections.  This lack
has not been remedied by follow-up work on thermodynamic depth, though
commentary at that time by Landauer \cite[p. 307]{Land88a} raised related
concerns.

Assuming one wants to use thermodynamic depth to measure complexity, Occam's
razor \cite{Occa64a} advises us to pick the simplest representation we can---in
this case, whichever selection of states gives the smallest depth; cf. Ref.
\cite[p. 193]{Page88a}. But this can always be trivially achieved by lumping
everything into one state, as just noted, which gives a vanishing depth.  More
confusingly there are even cases, as we'll see a bit later, where such lumping
is entirely appropriate.

Nor can the problem of state choice be reduced to that of coarse graining the
space of observables; as done in Ref. \cite[p. 194-5]{Page88a} and
elsewhere, for example in Refs. \cite{Zhan91a} and \cite{Foge92a}. While this
space can be readily represented by a finite alphabet, as done above---indeed,
digital measuring devices so represent it without even asking permission---the
problem is that the connection between what we measure and the underlying
process is often obscure to the point of total darkness.  (The definitions of
``measurement'' for Hamiltonian and quantum mechanical systems in Ref.
\cite{Page88a} shed no light on this point.)  It is certainly not desirable to
conflate a process's complexity with the complexity of whatever apparatus
connects the process to the variables we happen to have seized upon as handles.

One helpful step in developing any measure of complexity
is that it be calculated on simple illustrative examples which can be
thoroughly and unambiguously analyzed.  We now proceed to do this for a series
of examples---all of them based on Markov chains, if only to guarantee that
nothing especially tricky or esoteric is at issue.  In fact, we can interpret
each example as a type of one-dimensional spin-$1$ statistical mechanical
system; cf. Ref. \cite{DNCO}. (We emphasize that our results in other sections
are not restricted to this class of Markov processes.)

The hidden Markov models we analyze contain a set of ``internal'' states,
belonging to a finite alphabet ${\cal X}$, which are not
directly observable. At each time-step, there is some probability of
moving from the current state to any other, while ``emitting'' an observable
symbol drawn from another alphabet ${\cal A}$. We denote the probability of
going from internal state $i$ to internal state $j$ while emitting the
measurement value $s$ as $T^{(s)}_{ij}$. These models thus generate a pair
of linked stochastic processes, one over the internal states and the other
over the observable values, and only the latter is directly detectable.
Nonhidden Markov models are those where these two processes are one and
the same: where ${\cal A} = {\cal X}$ and $T^{(s)}_{ij} = 0$ unless $s = j$.

\begin{figure}[tbp]
\epsfxsize=2.7in
\begin{center}
\leavevmode
\epsffile{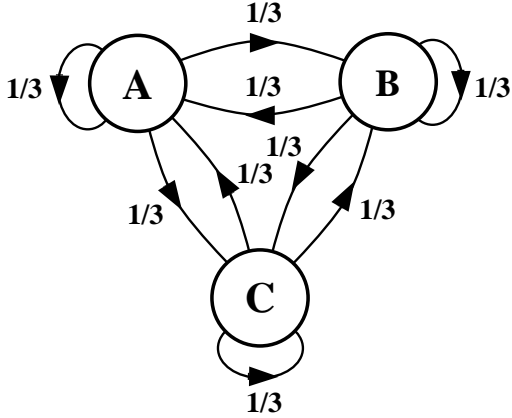}
\end{center}
\caption{A simple Markov chain that generates random sequences---${\bf BBAC}
\ldots$---with finite dive ($v = \log_2 3$) and so infinite total depth
(${\cal D}_L \rightarrow_{L \rightarrow \infty} \infty $).}
\label{shown-markov-model}
\end{figure}

Consider first a nonhidden system of three states ${\cal X} = {\cal A} = \{
{\bf A}, {\bf B}, {\bf C} \}$, each of which can go to any other, including
itself, with equal probability; see Fig. \ref{shown-markov-model}.  Here,
according to the prescription of Lloyd and Pagels, ${\cal D}_L = L \log_2 3$,
the total depth is infinite, and the dive is exactly equal to the entropy rate
of the observable sequences, i.e.  $v = \ObserveRate = \log_2 3$ bits per step.
The sequences generated are completely random, but neither the depth nor dive
vanish.

\begin{figure}
\epsfxsize=2.7in
\begin{center}
\leavevmode
\epsffile{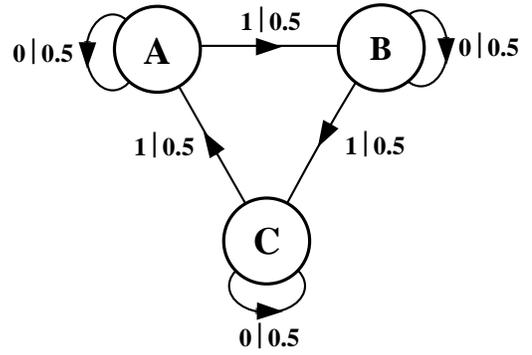}
\end{center}
\caption{A simple hidden Markov model that generates strings with finite dive
($v = \ObserveRate = 1$ bit per step) and infinite long-run depth. The edge
notation $s|p$ denotes that a transition is to be taken with probability $p$,
emitting measurement value $s$.}
\label{hidden-markov-model}
\end{figure}

Next, we hide the internal states $\cal X$ from observation, but at each time
step a measuring instrument emits one of two observable symbols $s \in {\cal A}
= \{ 0, 1\}$, as in Fig. \ref{hidden-markov-model}.  In this way we recover a
simple version of the micro-macroscopic distinction of Ref.  \cite{Page88a}.
The transition matrices $T^{(s)}_{ij}$ are, in this case,
\begin{eqnarray}
T^{(0)} & = & \left [ \matrix{ 1/2 & 0 & 0 \cr
							0 & 1/2 & 0 \cr
							0 & 0 & 1/2 \cr } \right ]
\end{eqnarray}
and
\begin{eqnarray}
T^{(1)} & = & \left [ \matrix{ 0 & 1/2 & 0 \cr
							0 & 0 & 1/2 \cr
							1/2 & 0 & 0 \cr } \right ] ~.
\end{eqnarray}
That is, each state either loops back on itself, emitting $s = 0$, or goes to
the next state in the chain, emitting $s = 1$, with equal probability.  The
dive, i.e. the entropy rate $\ObserveRate$ of the observables, is $v = 1$ bit
per step.  The entropy rate $\InternalRate$ of the internal states is also $1$
bit per step, since, given the current state, there are two possible,
equiprobable successors.  Moreover, while the system is a quite adequate source
of random sequences, macroscopic states $s \in {\cal A}$, as well as the three
hidden states, continue to deepen at the rate of $1$ bit per step.

Note that by inserting additional states between $\bf A$ and $\bf B$, that are
equally likely to either loop back to themselves on $s = 0$ or go to the next
state in the chain on $s = 1$, it is easy to go from Fig.
\ref{hidden-markov-model} to ``Rube Goldberg'' automata. These are
representations with elaborated sets of states with exactly the same
observable process and properties (i.e. with the same
${\rm Pr} ( \ldots , s_{-1} , s_0, s_1, \ldots)$, where $s_t \in \{0,1\}$),
but with increasing internal-state structure. Thus, there are inherent
ambiguities in using inappropriately baroque sets of states when
describing the structural properties of a process; ambiguities that
must be addressed somehow.

\begin{figure}
\epsfxsize=2.7in
\begin{center}
\leavevmode
\epsffile{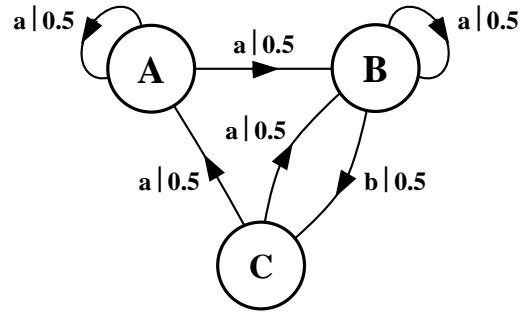}
\end{center}
\caption{A hidden Markov model of the logistic map symbolic
dynamics observed with a nongenerating partition.}
\label{unruly-machine}
\end{figure}

Finally, consider the symbolic dynamics of the logistic map of the unit
interval: $x_{t+1} = f(x_t) = 4 x_t (1 - x_t)$. Here the microscopic state
space is continuous: $x_t \in {\cal X} = [0,1]$, but we observe $x_t$ with a
binary-valued instrument ${\cal A} = \{ ``a" \sim [0,\hat{x}), ``b" \sim
[\hat{x},1] \}$, where $\hat{x}$ is the largest pre-image of $1/2$. When $x_t
\in [0,\hat{x})$ the instrument emits $s = a$ and when $x_t \in [\hat{x},1] $
it emits $s = b$.  This ``nongenerating'' partition of $\cal X$ leads to the
three hidden states that are coarse-grainings of $\cal X$: ${\bf A} \sim [0,
1-\hat{x})$, ${\bf B} \sim [1-\hat{x},\hat{x})$, and ${\bf C} \sim
[\hat{x},1]$.  Recalling that we can calculate the invariant distribution ${\rm
Pr} (x)$, the resulting stochastic finite-state model of the symbolic dynamics
process is shown in Fig. \ref{unruly-machine}.  (See Refs. \cite{Crut92c} and
\cite{Uppe97a} for more discussion of this example.)

The transition matrices for this process are
\begin{eqnarray}
{T}^{(b)} & = & \left [ \matrix{ 0 & 0 & 0 \cr
						0 & 0 & 1/2 \cr
						0 & 0 & 0 \cr } \right ]
\end{eqnarray}
and
\begin{eqnarray}
{T}^{(a)} & = & \left [ \matrix{ 1/2 & 1/2 & 0 \cr
						0 & 1/2 & 0 \cr
						1/2 & 1/2 & 0 \cr } \right ] ~.
\end{eqnarray}
The entropy rate $\InternalRate$, measured over the states, is $1$ bit per
step, but the dive $v = \ObserveRate$ (of the observables) is {\it lower}:
$v \approx 0.811$ bits per step. The states, in other words, are actually
worse---less predictable, deeper, and more demanding of memory (in a sense
made precise presently)---than the surface phenomena (sequences over $\cal A$)
they are supposed to explain. (Refs. \cite{Crut92c} and \cite{Crut91e} discuss
this curious phenomenon. A detailed mathematical analysis is found in
Ref. \cite{Uppe97a}.)

This example illustrates the measurement dependency of both randomness and
complexity. In contrast with the binary instrument just used, if the logistic
map is observed with a generating partition, for which infinite $a$-$b$
sequences are in correspondence with the microscopic states $x_t \in [0,1]$,
there is only a single internal state.  In this case, the internal state
entropy rate $\InternalRate$ is zero and the entropy rate of the observed
symbol sequences is $\ObserveRate = 1$ bit per symbol. It turns out that this
is the correct description of the logistic map dynamics; see Ref.
\cite{Devaney92} for an elementary exposition.

Readers will have already noticed, and been troubled by, the fact that all our
examples are simple sources of random strings, but have steep dives. According
to the definition, they are deep, complex processes, despite Lloyd and Pagels's
explicit statement that depth is small or vanishes for random processes.

\section{Causal States and $\epsilon$-Machines}

On the one hand, what these examples make clear is that we generally won't find
macroscopic states appropriate to measuring a process's statistical complexity
just by translating observables (via coarse-graining) into a finite alphabet.
On the other hand, especially in experimental work, we
often have no source of information other than the sequence of finite-precision
discrete-valued observables.  There is a fundamental difficulty here.
Moreover, part of the attraction of thermodynamic depth, compared to (say)
Kolmogorov-Chaitin complexity \cite{Kolm65,Chai66} and logical depth
\cite{Benn86}, was its claimed calculability from empirical data.

\begin{figure}
\epsfxsize=2.7in
\begin{center}
\leavevmode
\epsffile{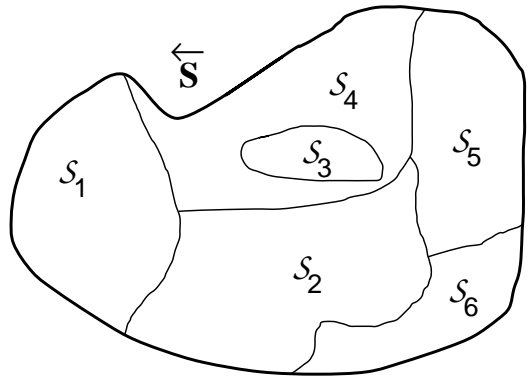}
\end{center}
\caption{A schematic representation of partitioning the set $\AllPasts$ of all
histories into causal states ${\cal S}_i$.  Within each
causal state all the individual histories $\past$ have the same conditional
distribution ${\rm Pr} (\Future | \past)$ for future observables.  Note that
the ${\cal S}_i$ need not form compact sets; we have simply drawn them that way
here for clarity.}
\label{epsilon-partition}
\end{figure}

There is at least one release from these ambiguities: it is found in the use
of {\it causal states}, as they are conceived of by {\it computational mechanics}---an
extension of statistical mechanics that explicitly accounts for a process's
structure \cite{Crut89,DNCO}. From the viewpoint of an observer, the idea is
that two trajectories leave one in the same causal state if they leave one
equally knowledgeable as to the future. More formally, a causal state $\cal S$
is an equivalence class over histories $\past$ of observed states, such that
all the sequences in the causal state give the same conditional distribution
for the semi-infinite future $\future$:
\begin{eqnarray}
\label{defcausalstate}
\epsilon(\past) & = & \{ {\past}^{\prime} ~|~ \forall ~ \future
~ {\rm Pr} (\future|{\past}^{\prime}) = {\rm Pr} (\future|\past) \} ~.
\end{eqnarray}
The causal-state equivalence classes form a partition of the set $\AllPasts$ of
all histories; see Fig. \ref{epsilon-partition}. Thus defined,
$\epsilon(\past)$ is a function from a history $\past$ to a set of histories,
which are the causal states ${\cal S}_i, i = 0, 1, 2, 3, \ldots$. We denote
the set $\{ {\cal S}_i \}$ of all causal states by $\CausalStateSet$. It
is convenient sometimes to have a function taking one from a history $\past$ to
the label $i$ of its equivalence class and, in a slight abuse of notation, we
will also call this $\epsilon(\past)$.

Since we need {\it some} choice of state if we are to apply depth at all, and
if we are not to consign it to the growing collection of subjective complexity
measures (see Ref. \cite{Feld97}), we might as well select a process's causal
states. What is notable, though, is that, while causal states were not designed
with this end in mind, they minimize dive.

The representation of a process consisting of the causal states and their
transitions is known as an $\epsilon$-{\it machine}. In the simplest setting, an
$\epsilon$-machine is a Markov chain over a finite number of causal states and
so can be compactly described by a labeled transition matrix $T^{(s)}_{ij}$,
notationally similar to that for the examples above. This matrix can be
calculated (analytically or empirically) from the distribution of observed
sequences, a procedure called $\epsilon$-machine {\it reconstruction}.

An $\epsilon$-machine lets us calculate the probability of different
sequences of observables.  It also leads to an invariant probability
distribution ${\rm Pr}({\cal S})$ over the causal states.  The resulting
complexity measure for a process is the statistical complexity $C_\mu$ which is
defined simply as the Shannon entropy of that distribution \cite{Crut89}:
$C_\mu = H[{\cal S}]$. $C_\mu$ measures the average amount of historical
information stored in the current state. Our results in section
\ref{optimality-results-for-epsilon} are not, however, restricted to cases
where the $\epsilon$-machine is finite Markovian, merely to ones
where there is a probability measure over the causal states.

A process's thermodynamic depth, and thus its dive, are defined with reference
to its macroscopic states, whatever we take those to be.  Due to the
ambiguities that follow from a prosaic interpretation of depth's
definition we propose to redefine depth, and by implication the dive, solely in
terms of a process's causal states. The first result of taking the
``macroscopic'' states to be the causal states is that the dive is the entropy
rate of the $\epsilon$-machine's internal-state process:
$v \equiv \InternalRate$, where ${\cal X} = \CausalStateSet$. The second result
is that by Eq. (\ref{v-bounded-by-entropy}) $v \leq C_\mu$. In fact,
$v < C_\mu$, if there is any mutual information in the observed sequences
$\stackrel{\leftrightarrow}{S}$, by Eq. (106) in Ref. \cite{DNCO}.

Causal state equivalence-classing guarantees that the $\epsilon$-machine is as
small as it can be and still be an accurate predictor of future observed
sequences; see subsection \ref{simplicity-of-epsilon} below. This makes
$\epsilon$-machines for both highly ordered and highly random sequences very
simple: a high degree of randomness means that many distinct sequences of
observables leave one equally uncertain about the future and, consequently,
those sequences all leave the system in the same causal state. In this way one
{\it derives} the desired ``boundary conditions'' for statistical complexity
measures---low for both simple and for random processes---from the underlying
principle of optimal prediction; that is, from Eq. (\ref{defcausalstate}).

\begin{figure}[tbp]
\epsfxsize=2.0in
\begin{center}
\leavevmode
\epsffile{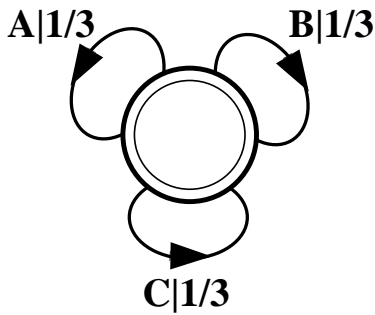}
\end{center}
\caption{The $\epsilon$-machine for the unhidden Markov model of
  Fig. \ref{shown-markov-model}. The internal entropy rate $\InternalRate$
  and the statistical complexity $C_\mu$ vanish since there is a single
  causal state.}
\label{EMachineFirstExample}
\end{figure}

These properties of causal states suffice to rescue the complexity analysis of
the examples from the confusions of the last section. The first
(Fig. \ref{shown-markov-model}) corresponds to an $\epsilon$-machine with a
single causal state ${\cal S}_0$ that returns to itself on three
separate, equally probable symbols ${\cal A} = \{ {\bf A}, {\bf B}, {\bf C} \}$.
(See Fig. \ref{EMachineFirstExample}.) The entropy rate $\ObserveRate$ of the
observed sequences is (as always)
preserved under the change of representation to causal states, but the entropy
rate $\CausalStateRate$ of the causal state process itself, i.e. the
now-redefined dive $v$, is, like the statistical complexity, zero.

\begin{figure}
\epsfxsize=2.3in
\begin{center}
\leavevmode
\epsffile{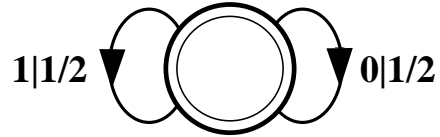}
\end{center}
\caption{The $\epsilon$-machine for the hidden Markov model of
  Fig. \ref{hidden-markov-model}. The internal entropy rate $\InternalRate$
  and the statistical complexity $C_\mu$ again vanish since there is a
  single causal state.}
\label{EMachineSecondExample}
\end{figure}

A similar fate awaits our second example (Fig. \ref{hidden-markov-model}).
Under causal state equivalence-classing, the three alleged states
collapse into one, yielding an ideal coin-tossing machine with a single
state and two transitions. (See Fig. \ref{EMachineSecondExample}.) Here
again the statistical complexity and the new dive vanish. Defining
depth in terms of a process's causal states leads us, in both
examples, to recover the intuitively correct notion that these sources of
purely random sequences are neither structurally complex nor store much
information about their history.

\begin{figure}
\epsfxsize=3.4in
\begin{center}
\leavevmode
\epsffile{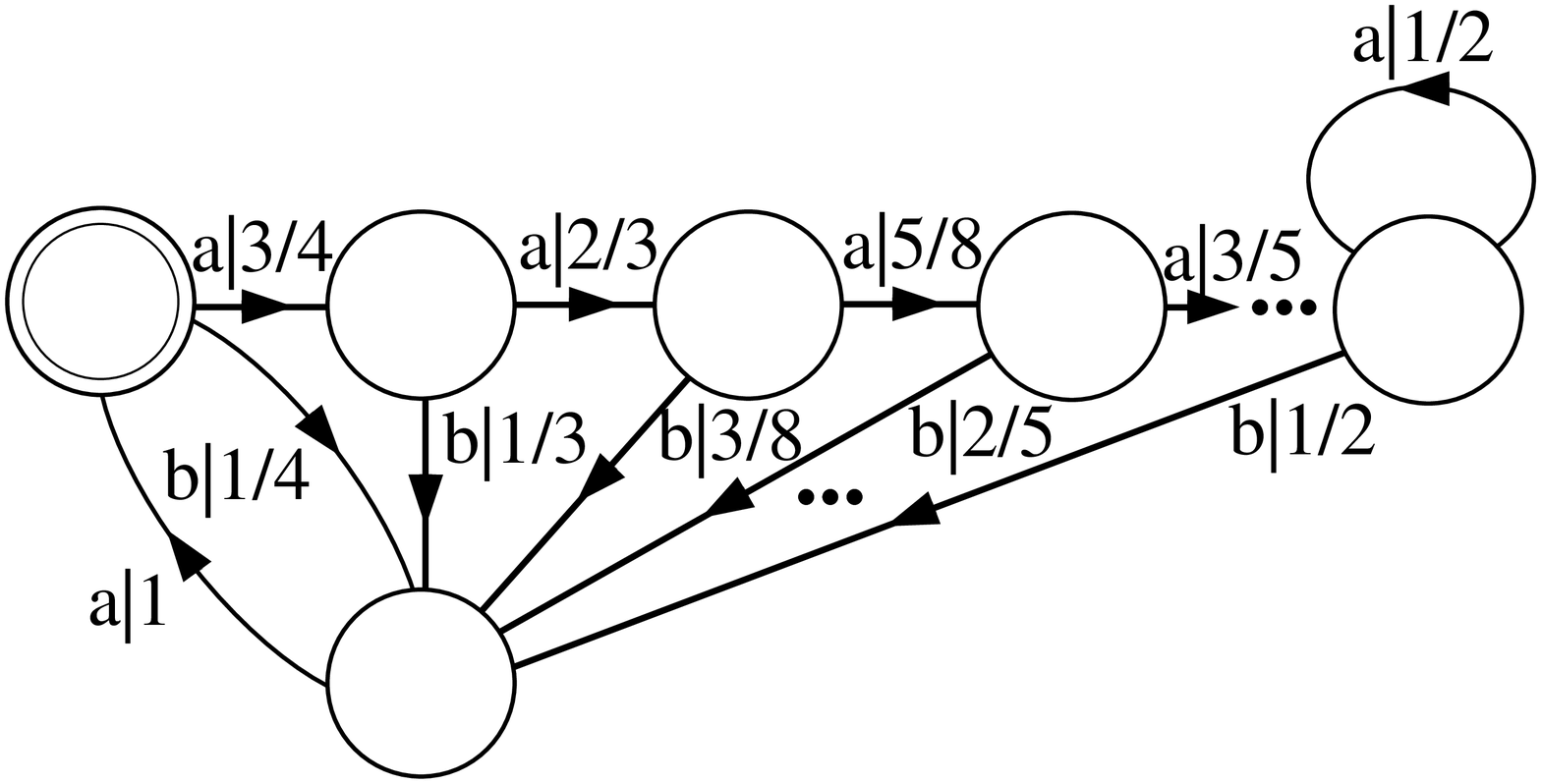}
\end{center}
\caption{The $\epsilon$-machine for the hidden Markov model of
  Fig. \ref{unruly-machine} has a countable infinity of causal states.
  The internal entropy rate $\InternalRate$ and statistical complexity
  $C_\mu$ are both positive, indicating that this is an intrinsically
  more complex process than the other two examples.
  }
\label{EMachineThirdExample}
\end{figure}

In our final example (Fig. \ref{unruly-machine}), the future conditional
distribution of observables depends only on how long it has been since the
last ``b'', leading to a countable infinity of causal states. (See Fig.
\ref{EMachineThirdExample}.) It turns out that the new dive and the statistical
complexity can be analytically calculated; one finds $v \approx 0.677867$ bits
per measurement and $C_\mu \approx 2.71147$ bits of historical memory are
stored by the process \cite{Crut92c,Uppe97a}. It is a more complex process
than the other two examples.

One of the desired properties of thermodynamic depth was that it accounted for
the history of the ``assembly process'' \cite[p. 187--9 and
{\it passim}]{Page88a}. We should emphasize that by definition causal states
account for a form of historical memory, though in an importantly different way.
Causal states measure the amount of historical information stored in a system.

\section{Optimal Shallowness of $\epsilon$-Machines}
\label{optimality-results-for-epsilon}

Working with the $\epsilon$-machine representation forces one to
distinguish between
\begin{enumerate}
\item sequences over coarse-grained observables $\cal A$,
\item sequences over causal states $\CausalStateSet$,
\item sequences over transitions, the labeled edges
$\{ (i,j,s) : T_{ij}^{(s)} > 0 \}$ ~.
\end{enumerate}
There is a
many-to-one relation between edge sequences and causal-state sequences and also
between edge sequences and observable sequences.  But, as we saw when we
defined the causal states as equivalence classes, Eq.  (\ref{defcausalstate}),
there is a function that takes a history to a causal state: namely,
$\epsilon: \; \AllPasts \; \mapsto \CausalStateSet$. One consequence is that one
can specify all of the relevant historical information by noting which of the
causal states the process is in, rather than recounting a possibly infinite
amount of information from the history $\Past$ that led to the current state.
That is, causal states provide a compression of a process's history.

\begin{figure}
\epsfxsize=2.7in
\begin{center}
\leavevmode
\epsffile{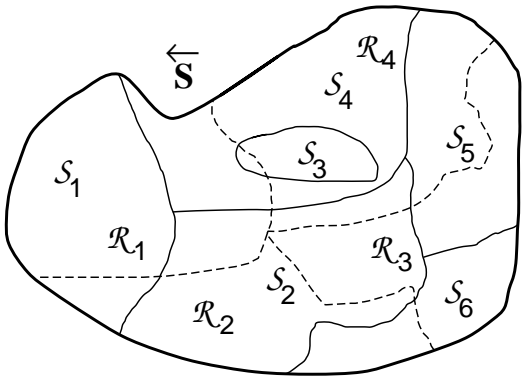}
\end{center}
\caption{An alternative set $\{ {\cal R}_i \}$ of states that
partition $\AllPasts$ overlaid on the causal states. (The ${\cal R}_i$ are
delineated by dashed lines.) The collection of all such alternative partitions
form Occam's ``pool''. Note again that the ${\cal R}_i$ need not be compact.}
\label{epsilon-and-bad-eta-partition}
\end{figure}

These distinctions and the historical compression are good motivations for
deciding which type of state to use for a process.  But these alone are not
enough, so let's consider alternatives to causal states, namely another set
$\AlternateStateSet$ of states, call them ${\cal R}$-states, that are determinable
from observed sequences and that, like causal states, partition $\AllPasts$;
see Fig. \ref{epsilon-and-bad-eta-partition}. We assume that these rivals to
the $\epsilon$-machine are, like the $\epsilon$-machine itself, restricted to
using only the past history of observables in their predictions, without any
other hints.

As one ranges over alternative choices of state---swimming around in Occam's
``pool'' of possible partitions---we will show that the $\epsilon$-machine
has a three-fold optimality: (i) no set of $\cal R$-states is more informative
about future observables than the causal states; of those choices of states
that are as predictive as the causal states, none has (ii) a smaller
statistical complexity nor (iii) a smaller entropy rate over the internal
states.  We conclude that none of the alternatives, if used to calculate the
depth, would give us a shallower dive than the causal states. We'll prove
these in order.

\subsection{Nothing Forecasts Better than an $\epsilon$-Machine}

Call the sequence of observables up to the present time $\Past$, the
random variable that is the next observable $S$, and the random variable that
is the whole sequence of future observables $\Future$. Recall that the function
$\epsilon: \; \AllPasts \; \mapsto \CausalStateSet$ returns the causal state the
$\epsilon$-machine is in after observing $\Past$ and define the function
$\eta: \; \AllPasts \; \mapsto \AlternateStateSet$ similarly for the
$\cal R$-states. We measure the forecasting ability of a set of states by
$H[\Future | {\cal R}]$,\footnote{This quantity can be infinite. In this case,
we should use $\lim_{L \rightarrow \infty} L^{-1} H[\Future^L | {\cal R}]$,
but for notational convenience this will be understood in the following.
The limit exists for all stationary processes \cite{Cove91}.} the
uncertainty that remains in the future observables once we know the current
state.  That is, the better the set of states is at forecasting---the more
prescient it is---the smaller this uncertainty. From Eq. (\ref{defcausalstate})
it follows that
\begin{equation}
{\rm Pr}(\Future|\epsilon(\Past)) = {\rm Pr}(\Future|\Past) ~,
\end{equation}
and so
\begin{equation}
H[\Future|\epsilon(\Past)] = H[\Future|\Past] ~.
\end{equation}
Since, for any random variables $X$ and $Y$ and function $f$,
\begin{equation}
H[Y|f(X)] \geq H[Y|X] ~,
\end{equation}
it follows that
\begin{eqnarray}
H[\Future | \eta(\Past)] & \geq & H[\Future | \Past] \\
                         & =    & H[\Future | \epsilon(\Past)]
\end{eqnarray}
and so
\begin{equation}
H[\Future|{\cal R}] \geq H[\Future|{\cal S}] ~.
\end{equation}
Thus, no alternative set $\AlternateStateSet$ of states sees the future better
than the causal states.

In what follows, we will put a hat over the name of any rival set of states
that is as predictive as the causal states, i.e. we refer to
states in $\hat{\AlternateStateSet}$ if and only if
$H[\Future|\hat{\cal R}] = H[\Future|{\cal S}]$.

\subsection{Nothing as Prescient as an $\epsilon$-Machine is Simpler}
\label{simplicity-of-epsilon}

\begin{figure}
\epsfxsize=2.7in
\begin{center}
\leavevmode
\epsffile{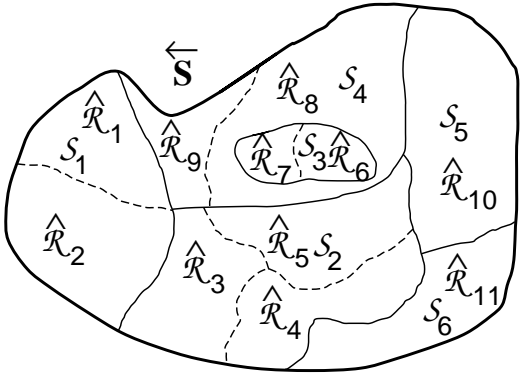}
\end{center}
\caption{Any alternative partition that is as prescient as
the causal states must be a refinement of the causal-state partition.  That is,
each $\hat{\cal R}_i$ must be a (possibly improper) subset of some ${\cal
S}_j$.  Otherwise, at least one $\hat{\cal R}_i$ would have to contain parts of
at least two causal states.  And so using this $\hat{\cal R}_i$ to predict the
future observables leads to more uncertainty about $\Future$ than using
the causal states.}
\label{refined-partition}
\end{figure}

Suppose we have a set $\hat{\AlternateStateSet}$ of states for which
$H[\Future|\hat{\cal R}] = H[\Future|{\cal S}]$.  Then, because the causal
states are equivalence classes with respect to future conditional
probabilities, the $\hat{\cal R}$-states must be refinements of these classes.
That is, rather than the situation depicted in
Fig. \ref{epsilon-and-bad-eta-partition}, we have the $\hat{\cal
R}$-partitioning shown in Fig. \ref{refined-partition}.  Otherwise at least one
$\hat{\cal R}_i$, considered as a set, would have to include histories that
belonged to at least two distinct causal states.  Such mixing of causal states
can only increase the uncertainty about the future sequence $\Future$ of
observables.  That is, for every $\hat{\cal R}_i$ there is a ${\cal S}_j$
such that $\hat{\cal R}_i \subseteq {\cal S}_j$ and so every causal state
is a union of $\hat{\cal R}$-states.

The result is that the causal state is a function of the $\hat{\cal R}$-state:
${\cal S} = g(\hat{\cal R})$. Thus,
\begin{equation}
H[{\cal S}] = H[g(\hat{\cal R})] \leq H[\hat{\cal R}] ~.
\end{equation}
But $H[{\cal S}]$ is $C_\mu$, the statistical complexity of the
$\epsilon$-machine, whereas $H[\hat{\cal R}]$ is the statistical complexity
of the $\eta$-machine---the set of $\hat{\cal R}$-states and their transitions.
Thus, of the optimally predictive alternative representations the
$\epsilon$-machine is the smallest, as measured by $C_\mu$.

An argument exactly parallel to the one in the preceding subsection shows, when
applied to the equally prescient alternatives, that
\begin{equation}
H[ \Future | \hat{\cal R} ] = H[ \Future | {\cal S} ] \Rightarrow
H[ \FutureBlock | \hat{\cal R} ] = H[ \FutureBlock | {\cal S} ] ~,
\label{MarginalizeFutureRate}
\end{equation}
for $L = 1, 2, \ldots$. (The opposite implication is not true, however.)
Thus, the causal states are also at least as informative about the next
(single) observable $S$ as any rival and, for that matter, about any finite
subsequence $\FutureBlock$ of the future. However, in the general case of the
previous paragraphs it is necessary to consider the whole semi-infinite future
because, potentially, coarser partitions can match these finite-$L$ predictive
powers. If, for instance, two histories have the same distribution for $S$,
but different distributions over the whole future, they belong to different
causal states. An $\hat{\cal R}$-state that combined those two causal states,
however, would enjoy the same ability to predict $S$ and its $\eta$-machine
would have a smaller statistical complexity.

\subsection{Nothing as Prescient as an $\epsilon$-Machine Has a 
Smaller Dive}

We will now show that the $\epsilon$-machine's dive ($v = \CausalStateRate$)
is at least as small as that of any equally prescient alternative. This also
turns on the fact that such $\hat{\cal R}$-states are refinements of the causal
states. The $\epsilon$-machine is deterministic in the sense of automata theory
\cite{DNCO}; that is, the present state ${\cal S}$ and the next observable $S$
together fix the next state ${\cal S}^\prime$, and so
$H[{\cal S}^\prime | S, {\cal S}] = 0$. Thus, we have
\begin{equation}
H[ S | {\cal S}] = H[{\cal S}^\prime , S | {\cal S}] ~.
\end{equation}
The $\hat{\cal R}$-machine, however, is not necessarily deterministic in this
sense, but all entropies are non-negative, so
$H[\hat{\cal R}^\prime | S, \hat{{\cal R}}] \geq 0$.
Since we are considering alternatives with the same predictive power as the
$\epsilon$-machine, i.e. alternatives for which $H[\Future | {\cal S}] =
H[\Future | \hat{\cal R}]$, then we have $H[S | {\cal S}] = H[S | \hat{\cal
R}]$.  On the one hand,
\begin{eqnarray}
H[\hat{\cal R}^{\prime}, S|\hat{\cal R}] & = & H[S|\hat{\cal R}] + H[\hat{\cal R}^{\prime}|S, \hat{\cal R}]\\
& \geq & H[S|\hat{\cal R}] \\
&   =  & H[S|{\cal S}] \\
&   =  & H[{\cal S}^\prime, S| {\cal S}] \\
&   =  & H[{\cal S}^\prime|{\cal S}] + H[S|{\cal S}^\prime,{\cal S}] ~.
\end{eqnarray}
On the other hand,
\begin{equation}
H[\hat{\cal R}^\prime, S|\hat{\cal R}]
   = H[\hat{\cal R}^\prime|\hat{\cal R}] + H[S|\hat{\cal R}^\prime,\hat{\cal R}] ~,
\end{equation}
as well, so
\begin{equation}
H[\hat{\cal R}^\prime|\hat{\cal R}] + H[S|\hat{\cal R}^\prime,\hat{\cal R}] \geq
	H[{\cal S}^\prime|{\cal S}] + H[S|{\cal S}^\prime,{\cal S}] ~,
\end{equation}
or
\begin{equation}
H[\hat{\cal R}^\prime|\hat{\cal R}] - H[{\cal S}^\prime|{\cal S}] \geq
	H[S|{\cal S}^\prime,{\cal S}] - H[S|\hat{\cal R}^\prime,\hat{\cal R}] ~.
\label{TransitionEntropyCompare}
\end{equation}
Since a causal state is a function of an $\hat{\cal R}$-state, the transition pair
$({\cal S}^\prime, {\cal S})$ is a function of the transition pair $(\hat{\cal
R}^\prime,\hat{\cal R})$, implying that $H[S|{\cal S}^\prime,{\cal S}] \geq
H[S|\hat{\cal R}^\prime,\hat{\cal R}]$. Thus, the RHS of Eq.
(\ref{TransitionEntropyCompare}) is non-negative and this implies that
\begin{equation}
H[\hat{\cal R}^\prime|\hat{\cal R}] \geq H[{\cal S}^\prime|{\cal S}] ~,
\end{equation}
which is the desired result; namely,
$v^{\hat{\AlternateStateSet}} \geq v^{\CausalStateSet}$.
That is, nothing which predicts as well as the $\epsilon$-machine has a
smaller dive than the $\epsilon$-machine does.

\section{Conclusion}

If one prefers processes over static descriptions and dislikes pretending
every natural thing is a
digital computer, thermodynamic depth seemed to be an attractive complexity
measure: ``one of the remarkably few thrusts in this area which is not
conspicuously vacuous,'' in the words of Landauer \cite{Land88a}.
Since total depth most likely
shares the incalculability, though not the formal uncomputability, of
Kolmogorov-Chaitin complexity and logical depth, it is not, at face value,
physically significant. Dive, the rate at which depth increases, is both
calculable and significant. We showed dive is the reverse-time Shannon entropy
rate of the stochastic process over the macroscopic states one takes the
system to be in. With nothing else said or added, however, depth typically
measures historical randomness; as do Kolmogorov-Chaitin complexity and the
Shannon entropy rate.

Unfortunately, Ref. \cite{Page88a}, which introduced depth, gave no clue as
to how macroscopic states are to be selected; though it strongly suggested
this is simply a matter of coarse-graining the space of microscopic states; cf.
\cite[pp. 194--5]{Page88a}. As we've shown, this approach produces manifestly
ambiguous results.

By way of fixing depth, we highlighted the key role of the choice of
macroscopic states.  The causal states of computational mechanics do not suffer
from the defects and ill-definedness that led to trouble with other sorts
of states.  The procedure that identifies them, $\epsilon$-machine
reconstruction, also gives us a way to calculate depth and dive. We removed
depth's ambiguities and recovered its claimed features by redefining it in
terms of the causal states.

We then gave our main results, showing that no alternative set of states to the
causal states contains more information about the future of observables.
Moreover, unless an alternative throws some of that information away it cannot
have a smaller statistical complexity or a lower dive.  Thus,
$\epsilon$-machines are optimally shallow.

\section*{Acknowledgments}

We thank David Feldman and Deirdre des Jardins for helpful discussions.  We
also appreciate comments on the manuscript by Dave Feldman, Jon Fetter, and
Nigel Snoad.  This work was supported at UC Berkeley by ONR grant
N00014-95-1-0524 and at the Santa Fe Institute by ONR grant N00014-95-1-0975
and by Sandia National Laboratory contract AU-4978.

\bibliography{spin}

\end{document}